\RequirePackage{fix-cm}
\documentclass[twocolumn,epjc3]{svjour3}  
\makeatletter
\providecommand{\@LN}[2]{}%
\makeatother
\smartqed  
\RequirePackage{graphicx}

\journalname{Eur. Phys. J. A}

\begin{document}\sloppy

\title{Total kinetic energy and mass yields from the fast neutron-induced fission of $^{239}$Pu
}

\author{Alexander Chemey\thanksref{addr1}
        Ashley Pica\thanksref{addr1}
        Liangyu Yao\thanksref{addr1}
        Walter Loveland\thanksref{addr1,e1}
        Hye Young Lee\thanksref{addr2}
                \and
        S.A. Kuvin\thanksref{addr2}
}

\thankstext{e1}{Corresponding author: lovelanw@onid.orst.edu}

\institute{Department of Chemistry, Oregon State University, Corvallis, OR, USA \label{addr1}
           \and
           P-27, Physics Division, Los Alamos National Lab, Los Alamos, New Mexico, USA \label{addr2}
}

\date{Received: date / Accepted: date}

\maketitle

\begin{abstract}
The total kinetic energy (TKE) release in fission is an important observable, constituting over 80\% of the energy released in fission (E$_{f}$ $\approx$ 200 MeV). While the TKE release in the $^{239}$Pu(n,f) reaction was previously measured up to 50 MeV incident neutron energy (E$_{n}$), there were features in TKE release at the highest values of E$_{n}$ that were puzzling. There was a marked flattening of TKE release from E$_{n}$ = 30 to 50 MeV, in disagreement with the clearly decreasing TKE observed from E$_{n}$ = 0.5 to 30 MeV. To verify and clarify this trend, TKE measurements at higher values of E$_n$ were made. We present absolute measurements of TKE release in $^{239}$Pu(n,f) from E$_{n}$ = 2.4 to 100 MeV. We used silicon PIN detectors to measure the fragment energies and deduce mass-yield curves using the 2E-method. We also discuss fission asymmetry and the relationships between approximate fission fragment mass and distortion.
\end{abstract}
\section{Introduction}
\label{intro}
Since the discovery of nuclear fission, and the great quantities of energy released in fission, the nature of the fission process and the division of its energy has been of great interest. It was determined that over 80\% of the $\sim$ 200 MeV energy release is in the form of fission fragment kinetic energy, while the remainder is divided between prompt gamma rays, neutrons, and radioactive decay of fission products.\cite{Un71}\cite{Va73}\cite{Vi85}\cite{Lo06} Resultingly, TKE is an important feature in understanding the fission process. This manuscript will refer to fission products and fission fragments separately, where fission fragments are the nuclei prior to prompt neutron emission and those immediately after as fission products.\cite{Ma06} 
\subsection{Energy release in fission}
\label{IApprox}
The total energy released in fission (E$_{f}$) is easily calculated from the excitation of the compound nucleus and the mass excesses of fission fragments, when they are known. This energy is released in several forms, and can be described as 
\begin{equation}
E_{f} = TKE + E_{pf} + E_{d}
\end{equation}
where E$_{pf}$ is the energy released in prompt deexcitation, and  E$_{d}$ is the energy released by fission product daughter nuclides due to deexcitations/relaxations, particle evaporations, and subsequent decays.\cite{Un71}\cite{Wa91} The sum of E$_{pf}$ and E$_{d}$ are therefore summarized as the excess energy available above the potential energy surface of the distorting nucleus.\cite{Ca17}\cite{Al20a}\cite{Goe13}\newline
\indent In the liquid drop model, a fissioning nucleus distorts along a long axis, balancing surface energy and Coulomb repulsion until a saddle deformation occurs.\cite{Bu19} The influence of static nuclear distortions on mass yields is significant, and recent time-dependent Hartree-Fock calculations indicate that the creation of asymmetric mass splits in thermal-neutron-induced fission of early actinides can be described without consideration of the Z = 50 magic number.\cite{Sc18a} This phenomenon is particularly influenced by shell gaps at N = 84, N = 88, Z = 52, and Z = 56 that are due to octupole deformation.\cite{Go19}\cite{Wi76} There is therefore a balance within asymmetric fission between more costly rest masses of statically-deformed fission fragments and the favored but more-deformed magic nuclei.\cite{Bu19}\cite{Al20b}\newline
\indent As the compound nucleus distorts through the saddle point, Coulomb repulsion repels the charge centers of proto-fragments beyond the surface tension and the nucleus splits at the scission point. \cite{Wa91}\cite{Moe70}\cite{Moe72}\cite{Moe01} As fission is driven by Coulomb repulsion, two useful approximations for TKE are derived from this relationship:
\begin{equation}
TKE (MeV) \approx Z_{1} Z_{2} e^{2}/(D (A_{1}^{1/3} + A_{2}^{1/3})
\end{equation}
and
\begin{equation}
TKE (MeV) \approx (0.1189 \pm 0.0011) Z^{2}/A^{1/3}+(7.3 \pm 1.5)
\end{equation}
where Z$_{n}$ and A$_{n}$ are the proton and mass number of the fission fragments and e$^{2}$ is the squared elementary charge,\cite{Lo06} 1.44 MeV-fm in Equation 2, and Z and A refer to the charge and mass number of the fissioning nucleus.\cite{Vi85}\newline
\indent Equation 2 is related to the Coulomb barrier (B$_{C}$), where D$_{0}$ (for compact spheres) is typically quoted as r$_0$ = 1.16 fm. In equation 2, D is a multiple of D$_{0}$, typically between 1.5 D$_{0}$ and 1.6 D$_{0}$.\cite{Ca15} While B$_{C}$ is valid for spherical point-charges, proto-fission fragments at the scission point are highly distorted, resulting in a larger distance between charge centers and a decrease in Coulomb repulsion. This parameterization assumes that TKE dependence is simply due to the separation of charge centers and their mutual repulsion. This is an oversimplification that neglects the velocity of the fission fragments as the nucleus distorts from compact shapes to the scission point. A recent TDHF analysis of the fission of $^{264}$Fm determined that pre-scission kinetic energy is responsible for approximately 22 MeV of the 274 MeV total fission energy.\cite{Si14} A more nuanced perspective on TKE would include the kinetic energy of heavy and light fragments at the scission point, in addition to the Coulomb repulsion at this point (see equation 11a in \cite{Bu19} for an example). As there is no observable that can be associated with this feature by the methods presented in this work, the kinetic energy at the scission point is neglected, and all kinetic energies are assumed to be due to Coulomb repulsion. This somewhat understates the ratio of D/D$_0$ of the compound nucleus at scission. \newline
\indent Equation 3 is an empirical relationship for the kinetic energy of fission derived from the Coulomb term Z$^{2}$/A$^{1/3}$.\cite{Vi85} The intercept term incorporates the kinetic energy of the fission fragments at the scission point. Both equations make a reasonable first-order approximation of TKE release.\newline
\indent Similar results regarding energy distribution can be obtained from the sum of single particle solutions of the Schr{\"o}dinger equation and the Strutinsky microscopic corrections to the liquid drop model.\cite{Br72}\cite{St68}\cite{Br81}\cite{Be84} The mean E$_{f}$ is approximately constant for the fission of a given compound nucleus, regardless of excitation energy, though the distribution of this energy may vary, as in equation 1.\cite{Un71} Given the large amounts of energy released in TKE during the fission process, measurements of TKE enable a detailed look at compound nucleus evolution towards the scission point.
\subsection{TKE trends}
\label{ITrend}
Increasing the excitation energy of the fissioning nucleus results in decreased TKE release as prompt nucleon-evaporation and fission fragment excitation increase. When multiple nucleon evaporations occur prior to fission, the compound nucleus excitation energy decreases. The TKE sharply increases above each subsequent n$^{th}$-neutron evaporation energy, resulting in important exceptions to the trend of decreasing TKE with increasing compound nucleus excitation.\cite{Pa19}\cite{Tu18}\cite{Ma20}\cite{Br86} The effect of n$^{th}$-chance reactions can be seen in parameterizations such as those of Lestone and Strother, who modeled these phenomena for E$_{n}$ $\leq$ 20 MeV.\cite{Le14} At the highest E$^{*}$, the compound nucleus rapidly distorts and ultimately splits with little regard for low-energy configurations, yielding increased fragment excitation and decreased TKE. The decrease in TKE is driven by the decrease in asymmetric TKE, in addition to the increased symmetric fission component (as symmetric TKE is lower than asymmetric TKE).\cite{Mue84}\cite{Na86}\cite{Ya18} An increase in TKE(E$_n$) with neutrons well above the known 4$^{th}$-chance fission threshold would therefore be unexpected.
\subsection{Prior work on fast neutron TKE}
\label{IPrior}
An early report by Unik and Gindler to Argonne National Laboratory in 1971 provided a comprehensive understanding of TKE to that time. There were assembled datapoints with thermal neutrons and fast neutron sources up to 14.1 MeV.\cite{Un71}\cite{Pr60} Other experiments have been conducted with deuteron- and proton-induced knockout reactions onto light targets (eg: $^{7}$Li(d,n)2$\alpha$)), which have nearly monoenergetic neutron spectra that vary with the incident particle energy.\cite{Cr33} Recent experiments have revisited the question of TKE as a function of excitation energy, due to applications with prompt criticality and in fast neutron and accelerator-driven fuel cycles.\cite{Ho71}\cite{Hi20}\cite{Br16} The fast neutron- and proton-induced fission TKE release has been recently measured for reactions with $^{232}$Th, $^{233-235,238}$U, $^{237}$Np, $^{239,242}$Pu, and $^{241}$Am targets.\cite{Ya18}\cite{Hi20}\cite{Ki17}\cite{Me16}\cite{Is08}\cite{Ru04} Recent evaluations of TKE for neutron-induced fission well above E$_n$ = 20 MeV have generally used the LANSCE/WNR spallation facility, which provides neutrons up to 200 MeV.\cite{Li06}\newline
\indent In the Unik report of 1971, there were four reports of TKE release for $^{239}$Pu(n$_{th}$,f), with measurements of pre-neutron-evaporation TKE ($_{pre}$TKE) varying from 174.4 to 179.3 MeV.\cite{Un71}\cite{Be67}\cite{Mi62}\cite{Ne66} A later study firmly fixed the pre-neutron TKE at 177.9 MeV, near the current value.\cite{Ma06}\cite{Re71} Later analyses of the differences between $^{239}$Pu(n$_{th}$,f) and $^{240}$Pu(sf) found that the $\sim$6.5 MeV excitation of the $^{240}$Pu nucleus from the former reaction resulted in a decrease of nearly 2 MeV in TKE.\cite{Wa84} Dependence of the TKE release in the reaction of $^{239}$Pu(n,f) with fast neutrons was first systematically studied when Akimov and collaborators reported data from from E$_{n}$ = 0 to 5.5 MeV.\cite{Ak71}\newline
\indent A more recent evaluation and extension of TKE(E$_{n}$) for $^{239}$Pu(n,f)was obtained in 2016, when an experiment at LANSCE/WNR measured TKE from E$_{n}$ = 0.5 to 50 MeV. That experiment observed a flattening in TKE at 32+ MeV,\cite{Me16} well above the 4$^{th}$-chance fission threshold. In light of this unexplained flattening, and aware of the influence that a thick target can have on TKE measurements,\cite{Goe17} a reevaluation of this reaction was undertaken.
\subsection{This report}
\label{herein}
\indent This article extends the data for TKE(E$_n$) past the E$_n$ = 50 MeV maximum. The TKE data reported here are benchmarked to the thermal-neutron-induced TKE of $^{239}$Pu using the same methods and instrumentation that measured fast neutron TKE, enabling absolute determinations of TKE. Additionally, mass yields and comparisons between TKE and distortion at the scission point are also reported for the first time for the fast neutron-induced fission of $^{239}$Pu.
\section{Methods}
\label{Meth}
\subsection{Facility and setup}
\label{MFacil}
This experiment was conducted at the Los Alamos National Lab (LANL) WNR facility, on Flight Path 15R (FP15R). “White spectrum” neutrons were produced by spallation of a thick tungsten target by 800 MeV protons, producing $\sim$ 10$^{5}$ – 10$^{6}$ n/s up to 100 MeV. The proton beam was structured in the form of 625 $\mu$s macropulses with a repetition rate of 100-120 Hz. Each macropulse was composed of 0.25 ns-wide micropulses separated by 1800 ns, enabling neutron time-of-flight energy measurements.\cite{Li06} Thick lead shielding collimated the neutrons down the flight path, focusing the beam to 1 cm at the end of the beampipe. The target was laser-aligned with the beam path aperture. The distance from spallation source to $^{239}$Pu target was measured to be 13.85(1) m. Irradiations were conducted over 7 days, with total data collection time of $\sim$ 160 hours. Figure 1 illustrates the geometry of LANSCE and the detector assembly.\begin{figure}
  \includegraphics[width=0.5\textwidth]{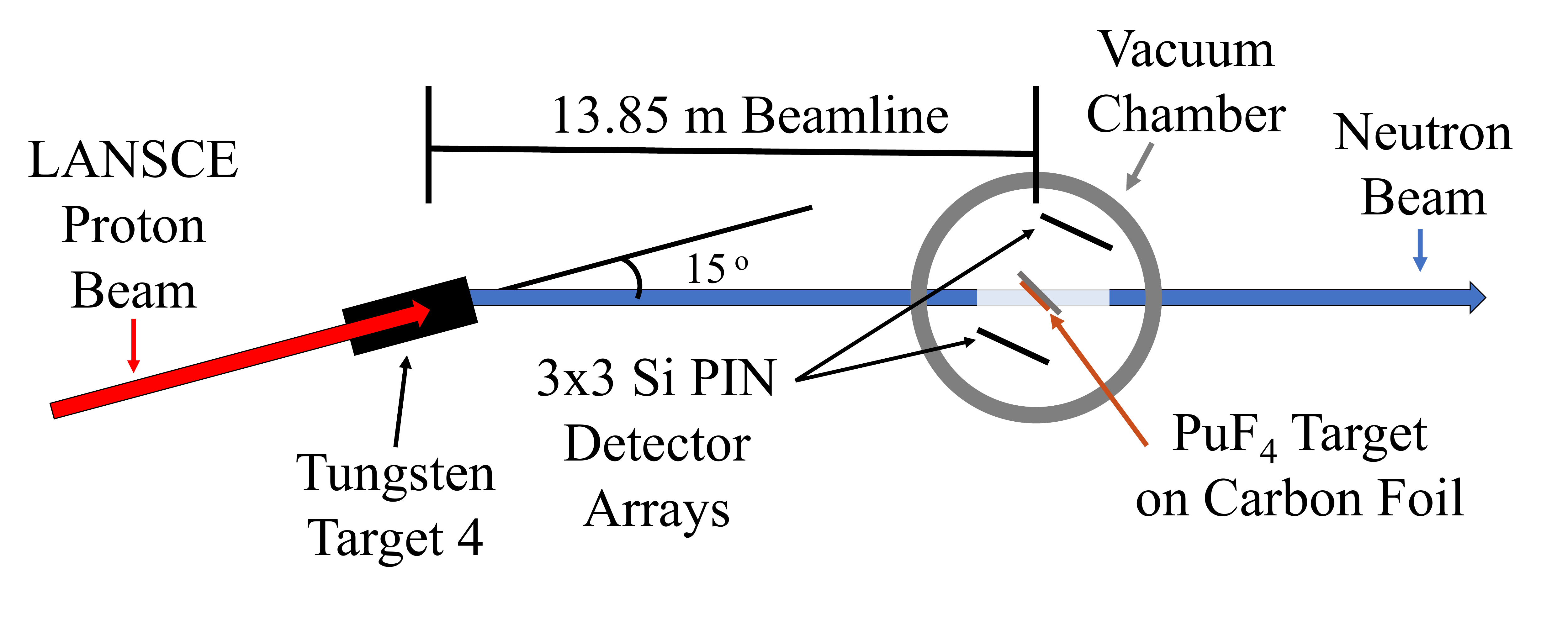}
\caption{The experimental setup at LANSCE/WNR FP15R.}
\label{fig:1}       
\end{figure}
\subsection{Target preparation}
\label{MTarg}
A 1 cm-radius circular target of 58.0 $\mu$g/cm$^{2}$ $^{239}$PuF$_{4}$ (determined by $\alpha$-counting) was deposited by the vapor deposition method onto a $\sim$100 $\mu$g/cm$^{2}$ carbon foil that was previously glued to an aluminum frame.\cite{Si20} The target assembly was then attached to the interior of an evacuated container for transport. After the scattering chamber was evacuated, the transport container was withdrawn and the target was exposed to the beam.
\subsection{Detection}
\label{MDet}
Twin arrays of nine 1 cm$^{2}$ Si PIN diodes (Hamamatsu S3590-09) arranged in a 3x3 square were located to the downstream left and upstream right of the neutron beam, 180$^\circ$ apart in the lab frame. The distance from the center of each detector array to the target was 3.5 cm. Each array covered $\sim$10\% of the solid angle of a half-sphere. Detection of coincident fission fragments was required to process signals by an analog AND gate.\newline
\indent The neutron energy was determined by time-of-flight measurements for every fission event. 4096 TDC channels spanned a timing window of 800 ns. The timing signal from the detector arrays was manually delayed at the beginning of the experiment to ensure that the photofission peak from the $^{239}$Pu($\gamma$,f) reaction appeared in the TDC spectrum. Determination of the neutron energy was by the relativistic time-of-flight (TOF) method, relative to the photofission peak.\cite{Si10} The uncertainty in the photofission peak mean was 0.24 ns, a relative uncertainty of 0.52\% on the 46.36 ns flight time.\newline
\indent Fission events that appear prior to the photofission peak are due to slow “wrap-around” neutrons with TOF greater than 1000 ns. The small amount of near-thermal neutrons incident on the target dominate the wrap-around regime due to the large $^{239}$Pu(n$_{th}$,f) cross-section ($\sigma_{f}$ = 748 b). Slow neutron fissions occur through the entire range of TDC signals, at the rate of 2.1(2) events/ns (measured in the wrap-around region). These wrap-around events are responsible for up to 90\% of the uncertainty in E$_{n}$, and result in an asymmetric uncertainty for any individual determination of E$_{n}$. Table 1 lists the asymmetric errors on E$_n$ at each of the ten bin mean neutron energies. At all neutron energies, the maximum uncertainty in measured E$_n$ is below 5\% when summed in quadrature.
\begin{table*}
\centering
\caption{Summarized data reported in this work, with asymmetric errors on E$_n$ at the mean values reported.}
\label{tab:1}  
\begin{tabular}{cccccc}
\hline\noalign{\smallskip}
E$_n$ Range(MeV) & Mean E$_n$ (MeV) & $_{pre}$TKE (MeV) & $_{post}$TKE (MeV) & $\sigma^{2}_{_{post}TKE}$ & Events \\
\noalign{\smallskip}\hline\noalign{\smallskip}
\relax[2.40-3.28] & 2.94$_{-0.13}^{+0.02}$ & 176.76(24) & 174.46(24) & 100(3) & 1697 \\
\relax[3.28-4.28] & 3.77$_{-0.17}^{+0.02}$ & 176.38(25) & 174.06(25) & 98(3) & 1653  \\
\relax[4.28-5.75] & 4.98$_{-0.21}^{+0.03}$ & 175.74(25) & 173.21(25) & 104(4) & 1653  \\
\relax[5.75-7.66] & 6.73$_{-0.27}^{+0.04}$ & 175.70(25) & 173.33(24) & 99(3) & 1653  \\
\relax[7.66-10.75] & 9.07$_{-0.33}^{+0.05}$ & 175.97(26) & 173.49(26) & 112(3) & 1652  \\
\relax[10.75-16.88] & 13.54$_{-0.44}^{+0.07}$ & 174.46(27) & 171.83(26) & 109(4) & 1648  \\
\relax[16.88-27.25] & 21.82$_{-0.56}^{+0.12}$ & 173.67(27) & 170.41(27) & 121(4) & 1649 \\
\relax[27.25-43.56] & 34.93$_{-0.68}^{+0.20}$ & 172.57(29) & 169.19(28) & 130(4) & 1650  \\
\relax[43.56-67.16] & 55.04$_{-0.78}^{+0.33}$ & 172.36(29) & 168.97(28) & 128(4) & 1659  \\
\relax[67.16-100.0] & 82.66$_{-0.89}^{+0.53}$ & 170.55(29) & 167.14(28) & 118(4) & 1533  \\
\noalign{\smallskip}\hline
\end{tabular}
\end{table*}
\subsection{Neutron evaporation correction}
\label{MNeut}
The detected ions are all post-neutron-evaporation, as prompt neutron evaporation occurs on a sub-femtosecond timescale, while the fragment time-of-flight is on the order of nanoseconds.\cite{Lo06} The compound nucleus before deexcitation offers the only absolute measurement of mass in the system. Determination of mass yields and correction for energy-loss through the target and foil require a reasonable approximation of the prompt-neutron evaporation both as a function of incident neutron energy and post-neutron fragment mass. The General Description of Fission Observables code (GEF 2019/1.2) is employed for this correction, calculating neutron multiplicities both as a function of incident neutron energy and post-neutron evaporation mass, $\nu_{n}$(E$_{n}$,A$_{post}$).\cite{Sc16} As noted previously for the high-energy fission of $^{235}$U(n,f), the GEF framework may systematically underestimate the total neutron evaporation by 5-10\% between 0 and 50 MeV.\cite{Ya18} Comparison of GEF neutron multiplicity calculations with data from literature assessments\cite{Hy64} found a similar underestimate for $^{239}$Pu(n,f) from E$_{n}$ = 0 to 15 MeV for GEF 2019/1.2. This is a sufficiently small deviation as to not meaningfully impact mass yields or fragment energy loss. Supplemental Figure 1 illustrates the effect of increasing E$_n$ on the fission product mass-dependent neutron multiplicity as modeled by GEF.\cite{Sc16}
\subsection{The double-energy method}
\label{M2E}
The double-energy (2E) method of fission product mass distributions is predicated on fission fragment emission in a momentum-conserving framework. Simultaneous detection of the kinetic energies of both fission products results in measurement of $_{post}$TKE. The mass of a fission product is inversely related to fractional contribution to TKE. As the 2E method measures post-evaporation products, an effort must be made to correct for neutron evaporation. By consideration of the neutron multiplicity, the pre-neutron mass yields (fission fragments) can be deduced.\cite{Si10} \newline
\indent The procedure to obtain the fragment kinetic energy is iterative, applying corrections for energy loss of fragments in the target and backing foils (which are mass- and charge-dependent) to then re-estimate fragment mass at the scission point. This continues until the differences in fragment mass between iterations is sufficiently small as to be nonexistent, defined here as an iterative mass difference of less than 0.1 amu.\cite{Si10} For most fission events, fewer than five iterations are required. It must be noted that mass resolution by this method is poor, with resolutions of only 4-5 amu, limiting the comparisons available between mass and TKE.\cite{Wa91}
\subsection{Charge distribution}
\label{MQ}
The treatment of charge yields assumed a uniform charge distribution (UCD) in fragments, where the charge distribution is of the same ratio as the compound nucleus, Z$_{1}$ = A$_{1}$Z$_{CN}$/A$_{CN}$, and Z$_{2}$ = Z$_{CN}$ – Z$_{1}$. The UCD assumption is reasonable for fission at high excitation energies, as shell effects "wash out,"\cite{Bi70} but there is some structure in the charge distribution at lower excitation energies due to proton shell effects.\cite{Ca15}\cite{Pa16} Simultaneous measurement of the mass-to-charge ratio in fission fragments requires the fragment velocities to be measured in addition to the energies, the 2v-2E method. That method is by far superior at resolving minor differences in TKE and fragment masses, though it suffers from poor solid angle coverage.\cite{Oe84}\newline
\indent While the UCD assumption does introduce a potential source of error due to incorrect fragment energy loss in foils, experiments probing the lack of uniform charge distribution have found that deviation from UCD for low excitation energies of the compound nucleus is less than a full proton per fission fragment.\cite{Ca15}\cite{Er79} Additionally, the energy lost in the foils is a small fraction of the kinetic energies measured, and the difference between two isobars that are a few protons apart is so small as to be negligible. Resultingly, the UCD assumption can be reasonably applied.
\subsection{Corrections and calibrations}
\label{MCorrCal}
The pulse-height defect of each detector was corrected for by the Schmitt method with a $^{252}$Cf source using updated fit parameters from the literature.\cite{Sc66}\cite{We86} Energy loss in the carbon backing and target were corrected for by using the Northcliffe-Schilling tables, assuming that the fission event took place in the middle of the target.\cite{No70} Due to the fast neutrons employed in this experiment, it was necessary to obtain a parameterization of the linear momentum transfer (LMT). Fractional LMT for the reaction $^{239}$Pu(n,f) was obtained from the literature and fit from E$_{n}$ = 7-120 MeV; with sub-7 MeV neutrons, LMT was assumed to be 100\%.\cite{He19}\cite{He20}
\subsection{Benchmarking}
\label{MBench}
The same target bombarded at LANSCE FP15R was taken to the Oregon State Triga Reactor (OSTR) and subjected to an intense (10$^7$-10$^8$ n/s) thermal neutron beam. The same equipment and data processing programs used to analyze the fast neutron results were also used to analyze the benchmark experiment. The areal density of the carbon foil (nominally 100 $\mu$g/cm$^{2}$) was systematically varied in the data analysis until experimental light and heavy kinetic energy curves of the front and back detector arrays overlapped, as only one fission product went through the carbon foil. From this analysis, the carbon foil thickness was determined, and applied to the data analysis as a part of the corrections and calibrations previously discussed. Measurement of effective thickness by scattering of the 6.118 MeV $\alpha$ from the decay of $^{252}$Cf resulted in a deduced thickness of 90 $\pm$ 10 $\mu$g/cm$^{2}$, in agreement with the deduced thickness from the thermal neutron data of 92.6 $\mu$g/cm$^{2}$. Data from the OSTR experiment agreed well with literature, with a measured $_{post}$TKE of 175.64(3) MeV, versus the fitting of Madland (175.55 MeV at 0 MeV) and the literature summary of 175.39(1) MeV for $^{239}$Pu(n$_{th}$,f).\cite{Ma06}\cite{Wa91}\newline
\indent To ensure that the fast neutron data are self-consistent with the benchmark experiment, coincident events from the wraparound region prior to the photopeak are used as an internal standard. The wraparound region is a mixture of sub-MeV neutron energies dominated by thermal-neutron-induced fission. The TKE for fission in the wraparound region should therefore agree well with the energy release obtained in the reactor experiment. A correction to measured fission product kinetic energies resulted in a wraparound region with a TKE of 175.6(5) MeV. The thermal-neutron-induced fission TKE spectrum is presented in Supplemental Figure 2, with wrap-around $_{post}$TKE from the fast neutron experiment as an inset.
\section{Results}
\label{Res}
\subsection{TKE released in fission vs E$_{n}$}
\label{RTKEEn}
The $_{pre}$TKE and $_{post}$TKE release from $^{239}$Pu(n,f) are presented in Table 1, with $_{post}$TKE depicted as a function of E$_{n}$ in Figure 2. Fission events were split into incident neutron energy bins that featured an approximately equal number of events. Data presented in these tables and plots are binned as to have approximately equal numbers of events. Ten neutron energy bins from E$_n$ = 2.4-100 MeV with counting errors of $\sim$2.5\% per bin were used. A plot of $_{post}$TKE for each neutron energy bin is available as Supplemental Figure 3.\newline
\indent With increasing E$_{n}$, the TKE release from $^{239}$Pu decreases. The data obtained here are best described by a log$_{10}$(E$_{n}$) fit for neutrons with E$_n$ greater than 1 MeV, TKE = (177.1 $\pm$ 1.0) - (4.9 $\pm$ 0.8) log$_{10}$(E$_{n}$) at a 95\% confidence interval with a standard deviation of the fit of 1.52 MeV. Figure 2 shows the binned data and its fit, as well as comparisons with to the literature up to E$_n$ = 50 MeV.\cite{Me16}\newline
\indent To complement binned data points as presented in Table 1 and Figure 2, the Gaussian means of rolling average TKE for every 0.1 MeV E$_{n}$ (with all events $\pm$ 5\% of E$_{n}$) is included in Figure 3, plotted to show the TKE as a continuous function of neutron energy. This method of plotting TKE is intended to highlight finer features than coarse binning allows, though poorer statistics at very high energies limit its utility to only the largest deviations. As can be seen in Figure 3, the fitline is appropriate for both the binned and rolling average TKE plots.
\begin{figure}
  \includegraphics[width=0.475\textwidth]{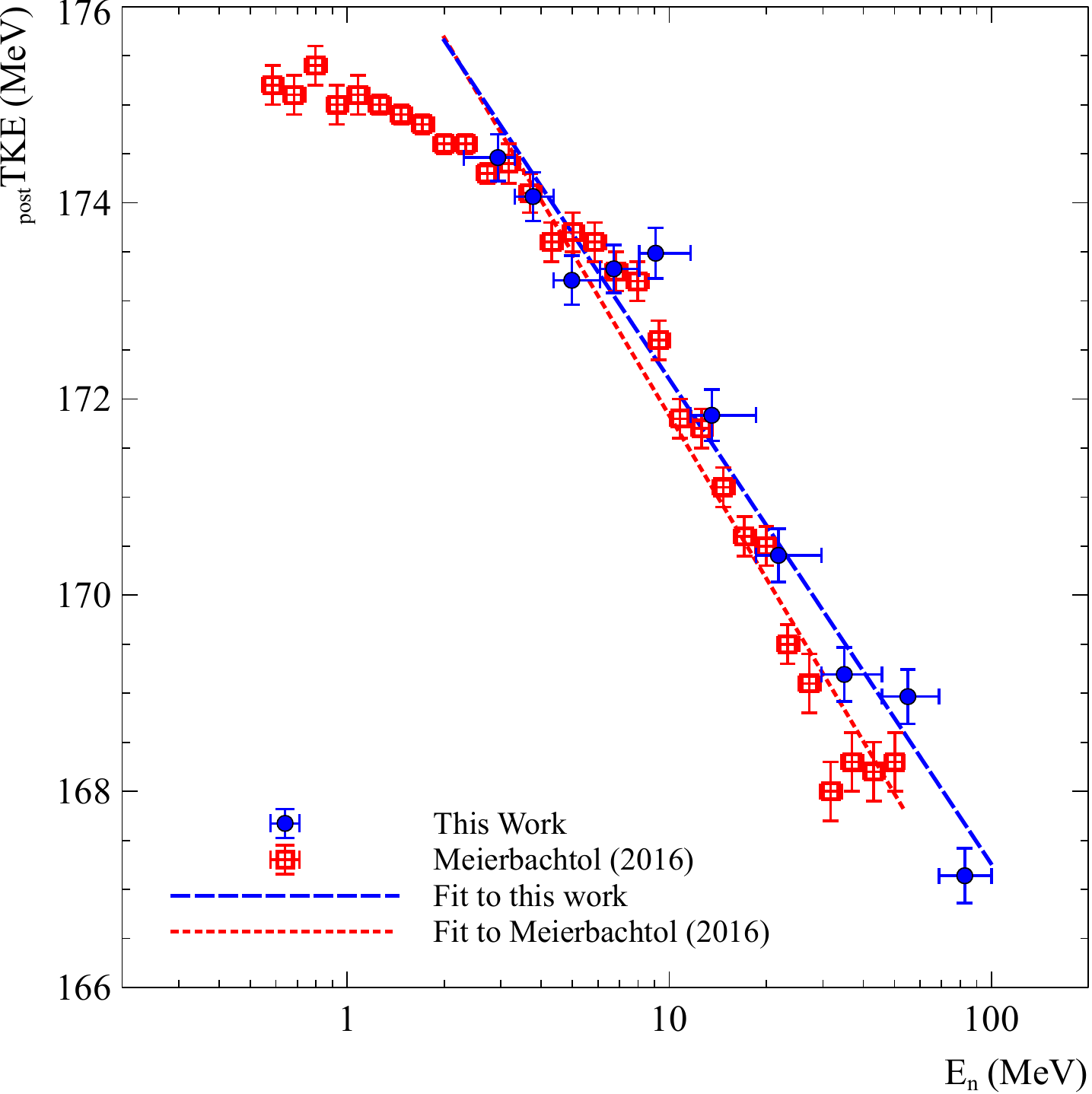}
\caption{Post-neutron TKE of $^{239}$Pu(n,f) for E$_{n}$ = 2.4 to 100 MeV (fit described in text), with comparison to literature\cite{Me16} and a fit to that work from 2.3 to 50.2 MeV. TKE error bars are statistical, E$_{n}$ error bars are the ranges included in each bin. Fit to literature: TKE = (177.3 $\pm$ 0.6) - (5.5 $\pm$ 0.6) log$_{10}$(E$_{n}$) at a 95\% confidence interval, with a standard deviation of the fit of 2.16 MeV.}
\label{fig:2}       
\end{figure}
\begin{figure}
  \includegraphics[width=0.475\textwidth]{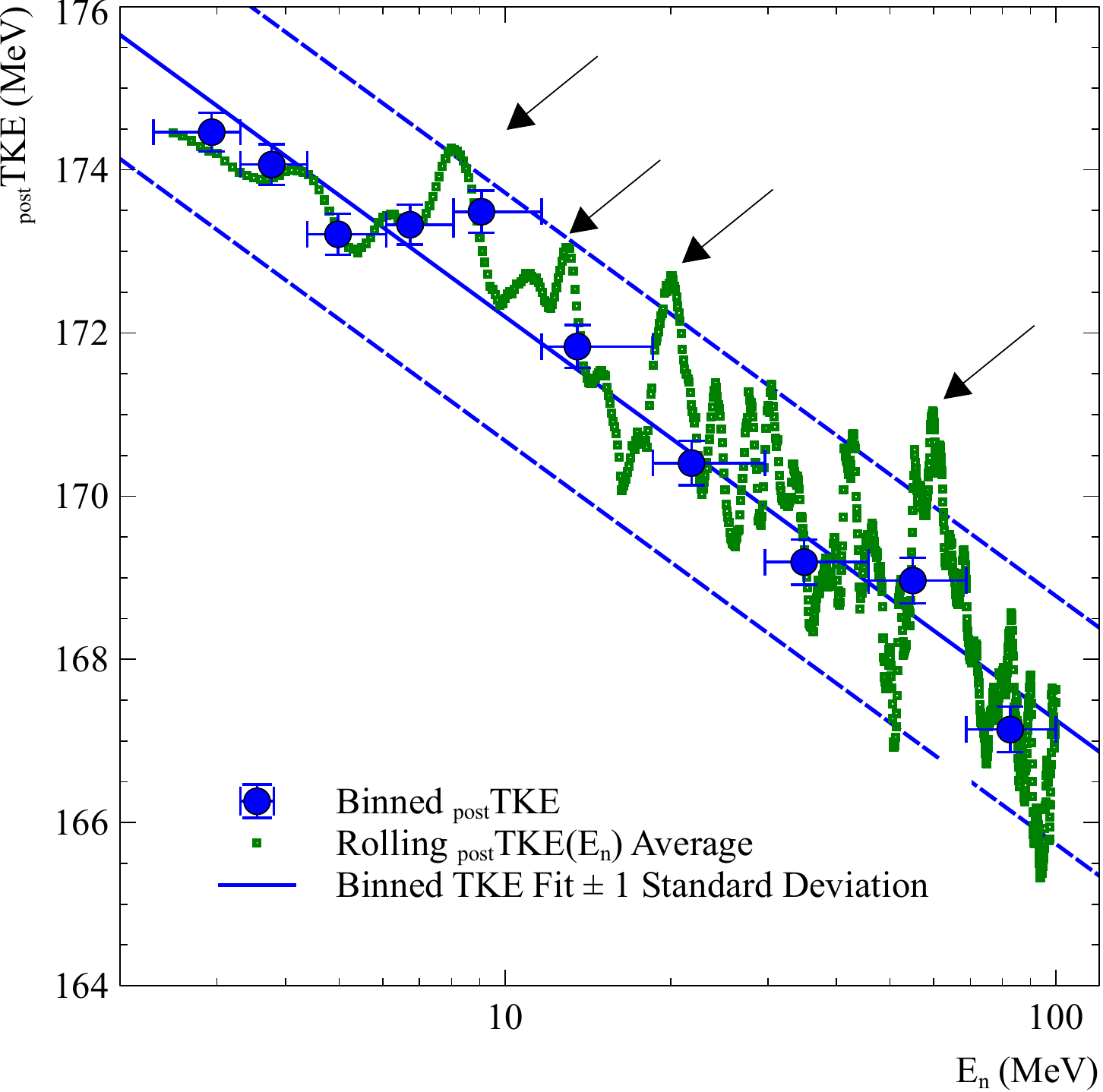}
\caption{Rolling average $_{post}$TKE plot on a $\pm$ 5\% basis, with comparison to binned data, including the fit to binned data $\pm$ 1 $\sigma$ of the fit. Arrows indicate deviations above 1 $\sigma$ that are highlighted in this work.}
\label{fig:3}       
\end{figure}
\subsection{Deviations from the TKE fit}
\label{RTKEDev}
Large and persistent deviations in in TKE(E$_{n}$) from the log$_{10}$(E$_n$) fit are of interest, as these may be indicative of structure in TKE(E$_{n}$). Deviations that show structure near the onset of n$^{th}$-chance fission are well understood. Multi-chance fission is possible when the incident neutron energy is greater than 5 MeV, steeply increasing to constitute a third of all fissions by 7 MeV.\cite{Le14} It is at these points in the TKE vs E$_{n}$ plots that a significant peak appears, well above the standard deviation of the fit. There is a close relationship between the onset of 2$^{nd}$-chance fission and the peak in the difference plot at 7-8 MeV (Supplemental Figure 4). Similarly significant peaks are also apparent at $\sim$ 12 MeV and $\sim$20 MeV, shortly after the onset of 3$^{rd}$- and 4$^{th}$-chance fission.\cite{Ch11} These observations are in accordance with parameterizations in the literature of TKE dependence for E$_{n}$ $\leq$ 20 MeV.\cite{Le14} Figure 3 also has arrows that indicate the approximate onset of 2$^{nd}$, 3$^{rd}$, and 4$^{th}$-chance fissions, and a significant deviation above 50 MeV that will be discussed later.
\subsection{$_{post}$TKE comparisons to literature and GEF calculations}
\label{RTKELitGEF}
As can be seen in Figure 2, the agreement between this experiment and previous experimental data is quite good between E$_n$ = 2 MeV and 7 MeV. Above this point, the general trend of decreasing TKE(E$_{n}$) is reproduced, though the literature\cite{Me16} TKE(E$_n$) (which continues up to 50.2 MeV) decreases in TKE more rapidly until E$_n$ $\sim$ 32 MeV, at which point the plateau in TKE is observed.\newline
\indent When comparisons are made between GEF predictions and experimental data for both $_{pre}$TKE and $_{post}$TKE, GEF appears to systematically overestimate TKE by $\sim$ 1\% across the data points evaluated, a rather good comparison (Figure 4 top). It is important to note that there is no GEF-predicted flattening in TKE(E$_{n}$) above E$_n$ = 32 MeV. Another minor point of deviation between GEF and the experiment is in the prediction of TKE variances. GEF systematically overestimates the variances on the TKE distribution by over 40\% for all incident neutron energies (Figure 4 bottom). This is similar to previous results obtained on TKE in actinide fission.\cite{Ya18}\cite{Ki17}
\begin{figure}
  \includegraphics[width=0.475\textwidth]{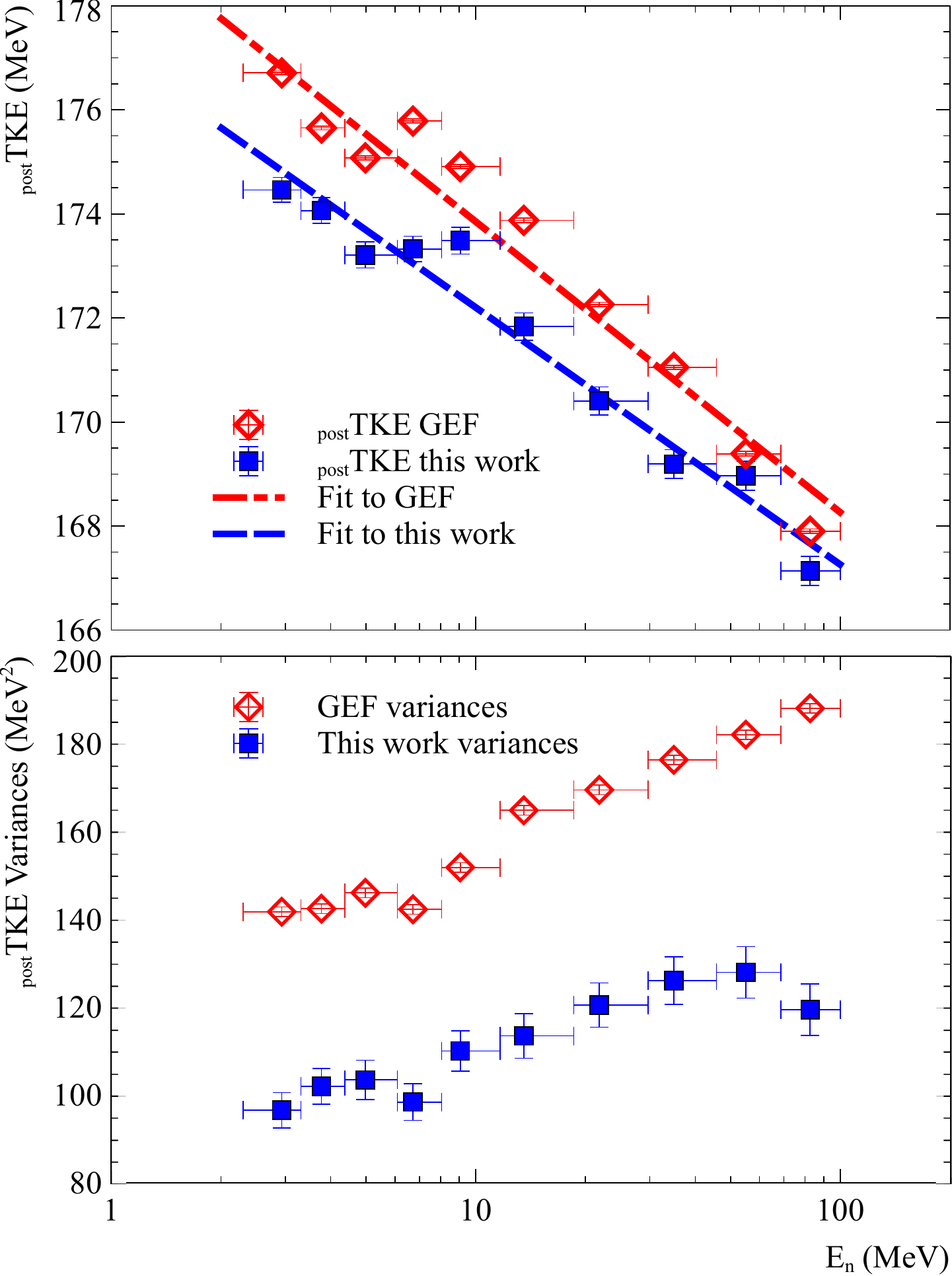}
\caption{Top: Plot of experimental $_{post}$TKE(E$_{n}$) vs GEF calculations of $_{post}$TKE(E$_{n}$) for $^{239}$Pu(n,f), with logarithmic fits to both datasets. The fit to GEF data is $_{post}$TKE = (179.4 $\pm$ 1.4) - (5.6 $\pm$ 1.2) log$_{10}$(E$_{n}$) at the 95\% confidence interval, with a standard deviation of the fit of 2.13 MeV. Bottom: Plot of experimental and GEF-predicted $_{post}$TKE variances and associated errors.}
\label{fig:4}       
\end{figure}
\subsection{Changing mass yield distributions with E$_{n}$}
\label{RMassEn}
Depicted in Figure 5 are the corresponding fission fragment distribution plots (normalized to 200\% yield) for the ten neutron energy bins from 2.4 to 100 MeV. (The thermal neutron fission fragment mass yields from the benchmarking experiment are in Supplemental Figure 5.) With increasing neutron energy, the contribution of symmetric fission increases, indicating the decreased importance of shell effects on fission fragments. In Table 2, we list the fitting details for mass yields (Supplemental Figure 6) from E$_{n}$ = 2.94 MeV to 82.66 MeV and peak-to-valley ratios.\newline
\indent Figure 6 is the rolling average mass distributions on a $\pm$10\% of E$_{n}$ basis. As indicated in this plot, the mass distributions in this reaction can be described generally as having three regimes: E$_{n}$ $<$ 10 MeV, 10 MeV $\leq$ E$_{n}$ $\leq$ 30 MeV, and E$_{n}$ $>$ 30 MeV. In the low energy regime below 10 MeV, the fission products are highly asymmetric, and there is little change in the bulk distribution of fission fragment masses with increasing En. In the intermediate energy regime, which extends to $\sim$30 MeV, the mass distribution steadily becomes more symmetric, though both symmetric and asymmetric fission modes are still distinguishable. Well above 30 MeV, symmetric fission simply dominates.
\begin{figure*}
\begin{center}
  \includegraphics[width=0.75\textwidth]{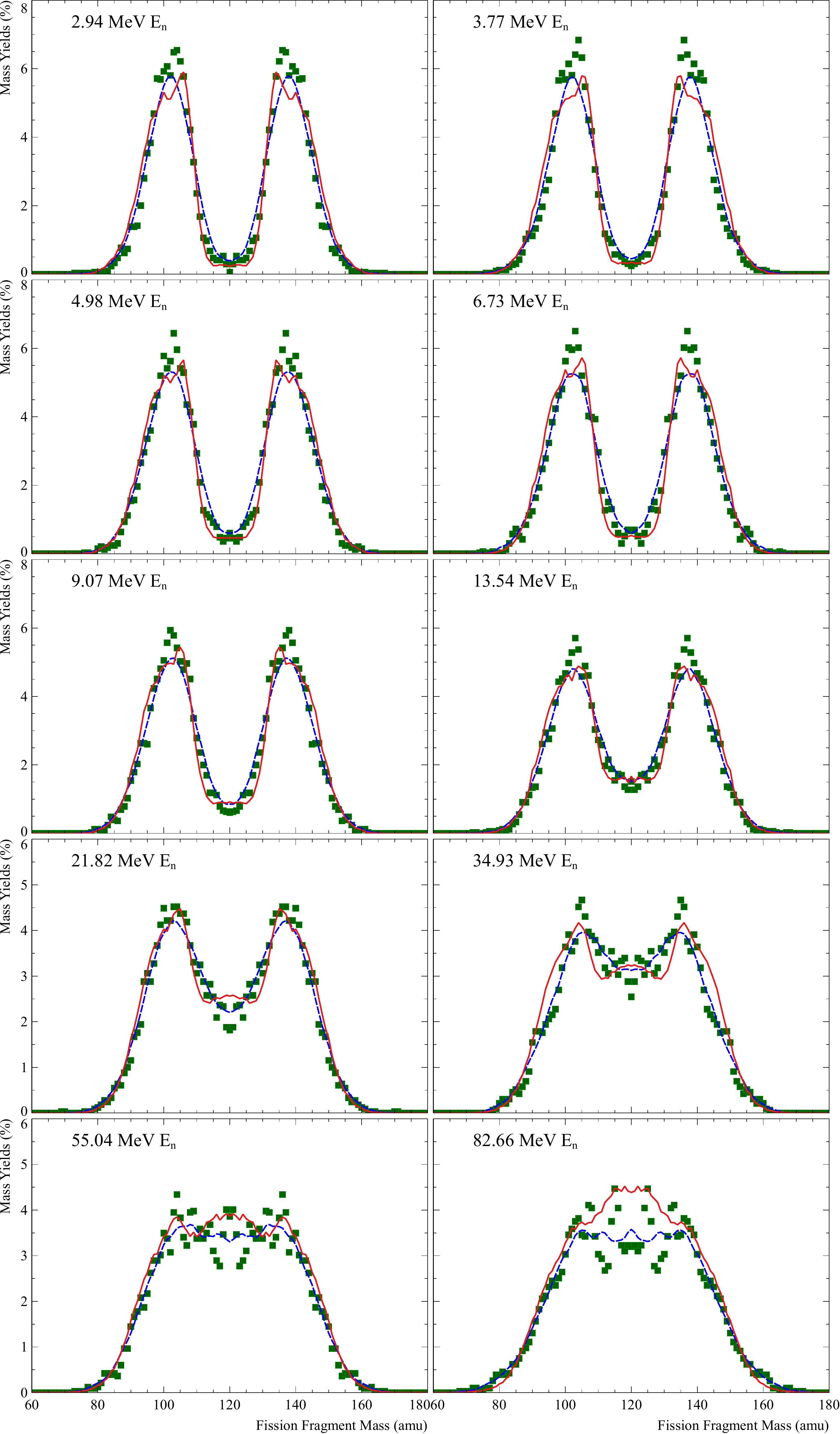}
\end{center}
\caption{Fission fragment mass distributions (normalized to 200\%) from this experiment as a function of incident neutron energy, compared with GEF-calculated mass yields and smoothed average ($\pm$ 5 amu) experimental mass yields. Key: Green squares, experimental mass yields; red line, GEF calculated mass yields; blue dashed line, smoothed experimental mass yields.}
\label{fig:5}       
\end{figure*}
\begin{figure*}
  \includegraphics[width=1\textwidth]{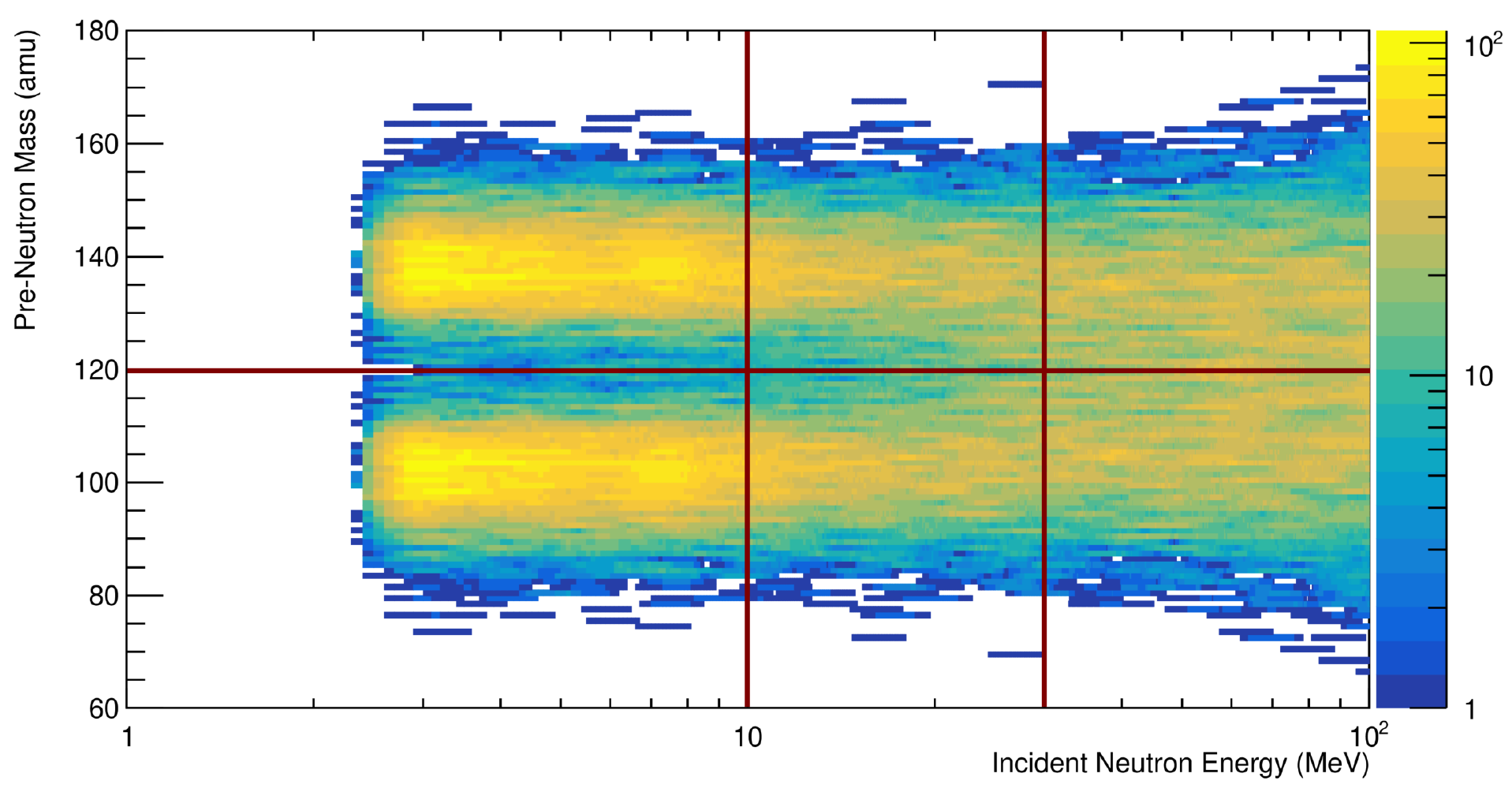}
\caption{Rolling average plot of fission fragment masses vs incident neutron energy from 2.4 to 100 MeV. The horizontal line is symmetric fission at A = 120, while vertical lines are located at E$_n$ = 10 and 30 MeV, indicating mass distribution regimes.}
\label{fig:6}       
\end{figure*}
\begin{figure*}
  \includegraphics[width=1\textwidth]{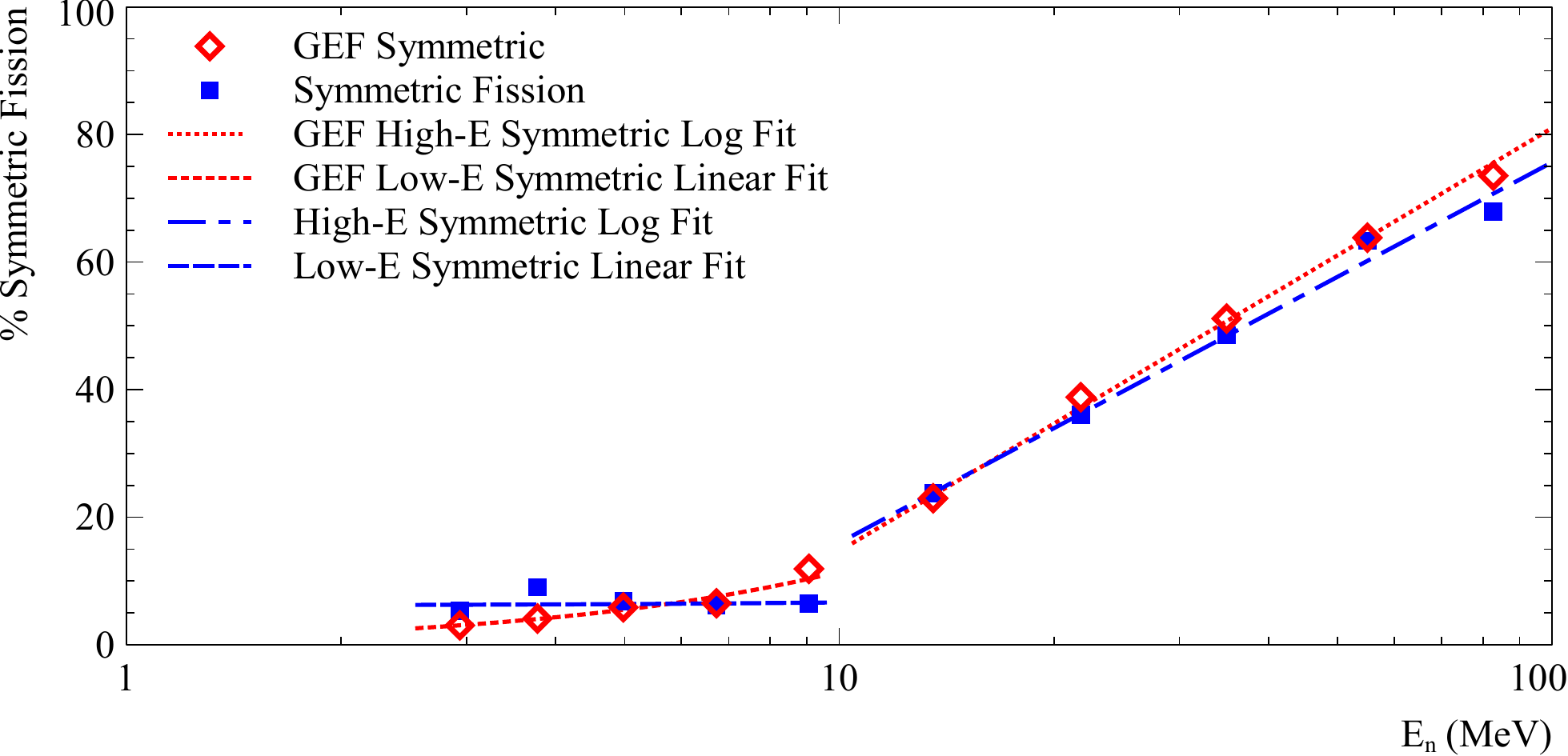}
\caption{Plot of pre-neutron symmetric and asymmetric path comparisons between the experiment and GEF calculations, with linear and logarithmic fits to E$_{n}$ discussed in the text.}
\label{fig:7}       
\end{figure*}
\begin{table*}
\centering
\caption{Fits to experimental mass yields. $\sigma$ is the Gaussian width of the asymmetric and symmetric peaks. \%Symmetric Contribution is determined by the peak heights and Gaussian widths. The Peak/Valley Ratio is the ratio of the maximum bin in the region of the asymmetric peak maximum and the bin at 120 amu.}
\label{tab:2}      
\begin{tabular}{cccccc}
\hline\noalign{\smallskip}
\multicolumn{1}{p{2.2cm}}{\centering E$_n$ Range (MeV)}
&
\multicolumn{1}{p{2.2cm}}{\centering \% Symmetric\break Contribution}
&
\multicolumn{1}{p{2.2cm}}{\centering Symmetric\break $\sigma$ (amu)}
&
\multicolumn{1}{p{2.2cm}}{\centering Asymmetric\break Peak (amu)}
&
\multicolumn{1}{p{2.2cm}}{\centering Asymmetric\break $\sigma$ (amu)}
&
\multicolumn{1}{p{2.2cm}}{\centering Peak/Valley\break Ratio}\\
\noalign{\smallskip}\hline\noalign{\smallskip}
\relax [2.40-3.28] & 6.0(7)\% & 50(12) & 138.33(11) & 5.72(10) & 120 \\
\relax [3.28-4.28] & 9.3(8)\% & 38(9) & 138.25(12) & 5.33(14) & 30.5 \\
\relax [4.28-5.75] & 7.4(1.9)\% & 46(32) & 138.24(14) & 6.42(17) & 23.0  \\
\relax [5.75-7.66] & 6.3(1.4)\% & 36(14) & 138.30(10) & 6.40(17) & 10.4  \\
\relax [7.66-10.75] & 6.5(1.2)\% & 23(5) & 137.88(20) & 6.8(2) & 10.0  \\
\relax [10.75-16.88] & 23.7(3)\% & 16.6(1.0) & 138.45(23) & 6.6(2) & 5.71 \\
\relax [16.88-27.25] & 35.9(4)\% & 16.4(1.0) & 138.1(4) & 7.3(4) & 2.57 \\
\relax [27.25-43.56] & 45.4(1.1)\% & 15.9(1.1) & 136.6(1.1) & 8.2(1.0) & 1.93  \\
\relax [43.56-67.16] & 62.2(5)\% & 15.3(8) & 139.2(7) & 7.3(6) & 1.25  \\
\relax [67.16-100.0] & 65.2(8)\% & 16.9(8) & 139.2(1.2) & 8.4(1.1) & 1.40  \\
\noalign{\smallskip}\hline
\end{tabular}
\end{table*}
\subsection{Comparison between experimental and GEF-calculated mass yields}
\label{RMassGEF}
\indent Comparisons with GEF pre-neutron mass yields indicate strong agreement with predicted shell effects, as seen in Figure 5. The mass yields obtained in this experiment are slightly narrower than the mass yields predicted by GEF, but in very good agreement overall.\newline
\indent The Gaussian fits to both asymmetric and symmetric components enable the areas underneath each curve to be determined. From this, comparisons to GEF predictions are possible, as well as a description of the trends in symmetric contribution to overall fission. As illustrated in Figure 7, until E$_{n}$ = 10 MeV, there is no statistically meaningful dependence of symmetric contribution as a function of E$_n$. For E$_n$ $>$ 10 MeV, the contribution of the symmetric path logarithmically increases. Above $\sim$30 MeV, the increase in symmetric fission decreases slightly, relative to GEF predictions. Overall, GEF replicates the trends observed very well. Fits to these ranges of E$_n$ are shown in Figure 7, with fit details in Table 3.
\begin{table}
\centering
\caption{Fits to Symmetric Fission as a function of E$_n$ for this experiment and GEF predictions. Errors are 1 $\sigma$. Low: E$_n$ = [0,10] MeV. High: E$_n$ = [10,100] MeV.}
\label{tab:3}   
\begin{tabular}{cc}
\hline\noalign{\smallskip}
Dataset \& E$_n$ Range & \% Symmetric(E$_n$) \\
\noalign{\smallskip}\hline\noalign{\smallskip}
Expt. Low & (6.1 $\pm$ 0.4) + (0.05 $\pm$ 0.06) E$_n$ \\
GEF Low & (-0.4 $\pm$ 0.3) + (1.19 $\pm$ 0.07) E$_n$ \\
Expt. High & (-44 $\pm$ 5) + (60 $\pm$ 4) log$_{10}$(E$_n$)  \\
GEF High & (-52 $\pm$ 5) + (66 $\pm$ 4) log$_{10}$(E$_n$) \\
\noalign{\smallskip}\hline
\end{tabular}
\end{table}
\subsection{Kinetic energy and fission asymmetry}
\label{KEFA}
By gating on events that feature asymmetric or symmetric fission fragments (defined as a heavy fragment mass within 5 amu of the asymmetric peak mean at $\sim$ 139 amu, or within 5 amu of symmetric fission at 120 amu), and then plotting the TKE of each pathway separately, the relationship between fission asymmetry and kinetic energy can be examined. As can be seen in Figure 8, there is no clear dependence on TKE for symmetric fissions, while asymmetric fission TKE decreases with increased E$_n$. This supports prior work that determined the decrease in TKE is not simply due to increasing contribution from the symmetric fission path, but that the TKE of the asymmetric fragments also decrease within E$_n$.\cite{Ki17} A logarithmic fit to the asymmetric fission path of $_{post}$TKE$_{asym}$ = (178.3 $\pm$ 1.7) - (4.4 $\pm$ 1.3) log$_{10}$(E$_n$) is obtained at the 95\% confidence interval, with a standard deviation of the fit of 2.39 MeV. The weighted average of all symmetric fission TKE was determined to be 167.2(2.1) MeV; as there are few symmetric fission events below E$_n$ = 10 MeV, determination of $_{post}$TKE$_{sym}$(E$_{n}$) is unreliable. For E$_{n}$ less than 10 MeV, $_{post}$TKE$_{asym}$ is approximately constant with a slight decrease up to 6 MeV followed by a slight increase after the 2$^{nd}$-chance fission threshold up to 10 MeV, while for E$_{n}$ above 10 MeV, $_{post}$TKE$_{asym}$ decreases more rapidly than the overall $_{post}$TKE. Parameters from all the fits reported here are summarized in Table 4, for ease of comparison.\newline
\indent To attribute variations in measured TKE between each neutron energy bin, estimates of TKE that assumed no difference in $_{post}$TKE$_{asym}$ from bin to bin were compared to estimates of TKE that assumed no difference in symmetric contribution from bin to bin. For E$_{n}$ $\leq$ 10 MeV, the changes in $_{post}$TKE were entirely due to changes in $_{post}$TKE$_{asym}$, with increased symmetric contribution not contributing to the difference in $_{post}$TKE. Above E$_{n}$ = 10 MeV, the estimated difference in mean TKE was approximately equal between increased symmetric fission and decreased $_{post}$TKE$_{asym}$, with both factors contributing at least 1/3$^{rd}$ of the changes in measured $_{post}$TKE. These results are summarized in Supplemental Figure 7.
\begin{figure}
  \includegraphics[width=0.5\textwidth]{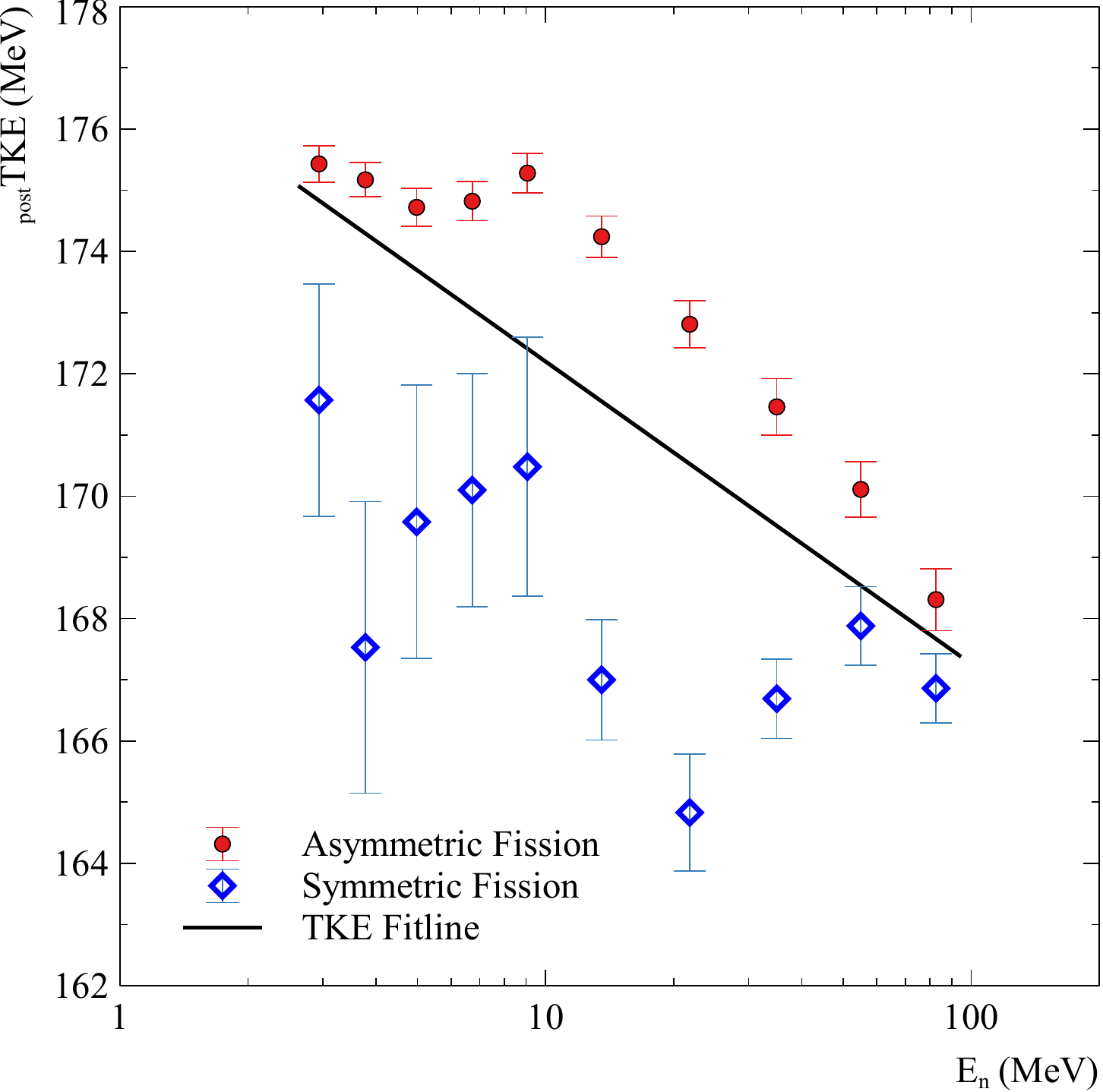}
\caption{$_{post}$TKE values gated on symmetric and asymmetric mass regions (see text for details). The overall fitline to $_{post}$TKE is shown for comparison. Error bars are uncertainty on measurements from the Gaussian fits.}
\label{fig:8}       
\end{figure}
\begin{table}
\centering
\caption{Summarized fits to TKE(E$_n$) from this work and the literature, with errors presented for the 95\% confidence interval.}
\label{tab:4}   
\begin{tabular}{cc}
\hline\noalign{\smallskip}
Fit Description & TKE Fit\\
\noalign{\smallskip}\hline\noalign{\smallskip}
$_{post}$TKE$_{thiswork}$ & (177.1 $\pm$ 1.0) - (4.9 $\pm$ 0.8) log$_{10}$(E$_{n}$)\\
$_{post}$TKE$_{[38]}$ & (177.3 $\pm$ 0.6) - (5.5 $\pm$ 0.6) log$_{10}$(E$_{n}$) \\
$_{post}$TKE$_{GEF}$ & (179.4 $\pm$ 1.4) - (5.6 $\pm$ 1.2) log$_{10}$(E$_{n}$)\\
$_{post}$TKE$_{asym}$ & (178.3 $\pm$ 1.7) - (4.4 $\pm$ 1.3) log$_{10}$(E$_n$)\\
\noalign{\smallskip}\hline
\end{tabular}
\end{table}
\subsection{Kinetic energy and fission fragment distortion at the scission point}
\label{RKEDDo}
When the TKE is measured and the fission fragment masses are calculated, reasonable determinations of the inter-nuclear distance can be produced by approximation of the charge to mass ratio (UCD assumption) and application of Equation 2. The ratio of distance between charge centers and the nominal contact distance between spherical nuclei (D$_0$) is typically expressed as D/D$_{0}$.\cite{Ca15} If D/D$_{0}$ $\approx$ 1.5, this indicates relatively mild distortion of the fission fragments at the scission point. Increased D/D$_{0}$ indicates greater distortion of fission fragments along the prolate axis, corresponding to excitation energy stored in the fission fragments. \newline
\indent When D/D$_{0}$ is plotted against fission fragment mass, there is clear differentiation between asymmetric and symmetric fission modes at low E$_{n}$. Mean values of D/D$_{0}$ are greater for symmetric fissions than asymmetric fissions at low-E$_n$, indicating greater distortion of symmetric fission fragments. For intermediate energies (10 MeV $\leq$ E$_{n}$ $\leq$ 30 MeV), the gap between symmetric and asymmetric fission distortion decreases slightly, and for high-E$_{n}$ events (E$_n$ $>$ 30 MeV), D/D$_0$ of fission fragment mass is far more constant. In this high E$_{n}$ region, D/D$_{0}$ is only slightly higher in the middle and slightly lower at the extremes, easily seen by the plotting of mean D/D$_{0}$(A) for the three energy regimes (Figure 9). It is important to note that these results imply increased distortion in asymmetric fission at high energies, with little variation in symmetric fission distortion as E$_n$ increases.
\begin{figure*}
  \includegraphics[width=1.05\textwidth]{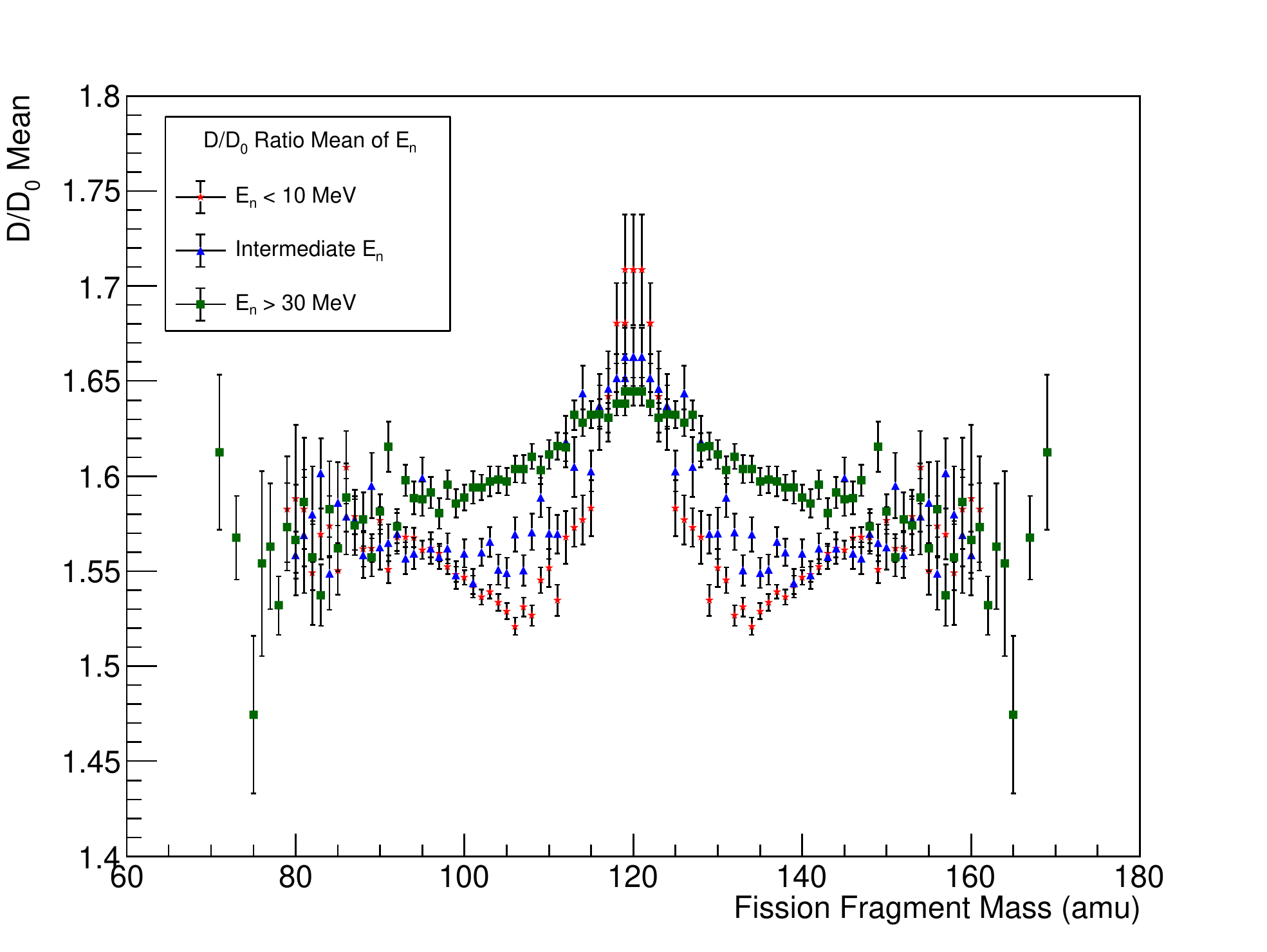}
\caption{Mean D/D$_{0}$ as a function of fission fragment masses, separated by E$_{n}$ regime.}
\label{fig:9}       
\end{figure*}
\section{Discussion}\label{Disc}
\subsection{The flattening TKE above 30 MeV}\label{FlatTKE}
A hypothesis of this experiment was that, far above the 4$^{th}$-chance fission threshold, there would be a consistent decrease in the dependence of TKE on E$_n$. The TKE measured at E$_n$ $\geq$ 50 MeV are unexpectedly high relative to the trend observed from data obtained at lower energies (see Figure 2 and Table 1). While not indisputable evidence for the flattening previously observed above 30 MeV,\cite{Me16} the observed plateau is suggestive of further unknown factors in these experiments. Although the downward trend in TKE does return in the final E$_n$ bin, this increase in TKE from $\sim$30 MeV to $\sim$70 MeV may be more significant than simple binning could explain. Further evidence can be observed in the rolling average $_{post}$TKE(E$_{n}$) plot of Figure 3. The fitline residuals are dramatically worse for the penultimate 55.04 MeV bin, with a more significant peak apparent from 55-70 MeV than those identified as due to the 2$^{nd}$, 3$^{rd}$, and 4$^{th}$-chance fission.\newline
\indent Careful reexamination of literature TKE reports for experiments that are conducted at LANSCE suggest that a similar flattening is observed for the reactions $^{232}$Th(n,f), $^{235}$U(n,f), in this report, and in the literature study of $^{239}$Pu(n,f).\cite{Ya18}\cite{Ki17}\cite{Me16} It looks increasingly likely that this increase in TKE is some form of signal rather than noise.
\subsection{Potential explanations}
\label{Explanations}
The question must therefore be asked: What could be causing this increase in apparent $_{post}$TKE vs the trends observed until these final data points?\newline
\indent 1)	Fission and deexcitation widths may see a large change at the highest excitation energies, resulting in a different distribution of E$_f$ above this energy. \newline
\indent 2)	There may be a competing reaction that has not been considered at the LANSCE facility. Equation 3 would suggest that the TKE released in the reaction $^{239}$Pu(p,f) would be at least 3 MeV greater than the comparable neutron-induced fission.\cite{Vi85} First-order calculations suggest that the slowest protons to overcome the\break Coulomb barrier would indeed arrive at the same time as neutrons in the $\sim$30 MeV range, though this is dependent on the distance from the spallation source and atmospheric conditions. To address this, MCNP\cite{MCNP} calculations of both W(p,p') from the spallation target and (n,p) reactions from materials in the beamline and in the scattering chamber were sought. Due to braking in air, protons leaving the spallation target with less than approximately 50 MeV energy are unlikely to contribute much to the overall TKE. It is estimated that approximately 4\% of the particles produced with greater than 50 MeV energy are due to protons.\cite{Za20} It is therefore unlikely that any significant effect on TKE would be observed.\newline
\indent 3)	There could be new physics at or approaching the scission point.
\section{Conclusion}
\label{Conc}
A thin homogenous vapor-deposited target of $^{239}$PuF$_{4}$ was used with semiconductor detectors to measure kinetic energies of fission products. The TKE-dependence of the reaction $^{239}$Pu(n,f) was measured for E$_{n}$ 2.4 to 100 MeV and correlated to fission fragment masses, fission paths, and distortion of the compound nucleus. The data obtained are absolute measures of TKE, as they are internally benchmarked to the well-known TKE release of $^{239}$Pu(n$_{th}$,f). A peculiar flattening in TKE(E$_n$) is confirmed in this experiment, apparent only above 30 MeV. Future work should study TKE trends in charged ion-induced fission of excited nuclei, both by knockout reactions and fusion-fission mechanisms. Special effort must be made to correlate the plateau in TKE with other observables, such as prompt gamma ray deexcitation or prompt particle emission. 
\section{Acknowledgements}
\label{Acknowl}
\indent We are grateful for the support of the facility operators at LANSCE and the OSTR for providing stable sources of neutrons. We are similarly grateful for the work of R. Yanez who designed the container used for shipping targets, C. Prokop for assisting with target handling at LANSCE prior to the experiment, and N. Fotiades for many fruitful conversations.
\indent This material is based upon work supported in part by the U.S. Department of Energy, Office of Science, Office of Nuclear Physics under award number DE-FG-06-97ER41026 (OSU) and under contract\linebreak 89233218CNA000001 (LANL). University collaborators acknowledge support from this work from the DOE-NNSA Stewardship Science Academic Alliances Program under Grant No. DE-NA0003907. This research benefited from the use of the LANSCE accelerator facility. Nuclides used in this research were supplied by the United States Department of Energy Office of Science by the Isotope Program in the Office of Nuclear Physics.
\section{Author contribution statement}
\label{contrib}
The experiment was designed and proposed and overseen by WL. Experimental data were collected by AC, AP, LY, and WL with critical assistance from HYL and SAK. AC and WL performed the primary data analysis. All authors contributed to the manuscript preparation, and have read and approved the final manuscript.
\section{Statement of data availability}
\label{Avail}
The datasets and analysis programs employed are available from the corresponding author on reasonable request. Supplementary material can be found online at the journal website.

\end{document}